# VISUALLY IMAGE ENCRYPTION AND COMPRESSION USING A CNN BASED AUTO ENCODER

Mahdi Madani and El-Bay Bourennane

ImViA Laboratory(EA 7535), Université Bourgogne Europe,21000 Dijon, France

## ABSTRACT

*This paper proposes a visual encryption method to ensure the confidentiality of digital images. The model used is based on an autoencoder using aConvolutional Neural Network (CNN) to ensure the protection of the user data on both the sender side (encryption process) and the receiver side(decryption process)in a symmetric mode. To train and test the model, we used the MNIST and CIFAR-10 datasets. Our focus lies in generating an encrypted dataset by combining the original dataset with a random mask. Then, a convolutional autoencoder in the masked dataset will be designed and trained to learn essential image features in a reduced-dimensional latent space and reconstruct the image from this space. The used mask can be considered as a secret key known in standard cryptographic algorithms which allows the receiver of the masked data to recover the plain data. The implementation of this proposed encryption model demonstrates efficacy in preserving data confidentiality and integrity while reducing the dimensionality (for example we pass from 3072 Bytes to 1024 Bytes for CIFAR-10 images). Experimental results show that the used CNN exhibits a proficient encryption and decryption process on the MNIST dataset, and a proficient encryption and acceptable decryption process on the CIFAR-10 dataset.*

## 1. INTRODUCTION

In our days, billions of digital data transit over the different networks existing over the world such as Internet, mobile networks, networks of connected objects, satellites, etc. An important part of these networks uses a wireless channel for communications where it is difficult to prohibit physical access to transmitted data. Therefore, guaranteeing the security and confidentiality of user information has become difficult, but a more than obligatory task. To meet these data protection requirements, several techniques have been explored for years suchas standard encryption algorithms (stream ciphers and block ciphers) used to protect sensitive data against unauthorized access and interception, hash functions used to ensure the integrity of sensitive messages and detect any change caused by transmission errors (channel, source, etc.) or by an attack. In the realm of image encryption, the quest for robust ciphering techniques has led to the exploration of traditional cryptographic methods, like chaotic systems used as pseudo number generators [1], Advanced Encryption Standard (AES) algorithms adopted in various domains [2, 3, 4], including secure communication protocols, password encryption in Wi-Fi networks, and data compression software. In the last decade, Artificial Intelligence (AI) and especially Convolutional Neural Network (CNN)-based models have emerged as powerful tools in the domain of computer vision, object detection, and image analysis and processing [5], leveraging hierarchical feature extraction through convolutional, pooling layers, activation

International Journal of Computer Networks & Communications (IJCNC) Vol.17, No.2, March 2025

functions, CNNs excel in discerning intricate patterns and representations within image data, in the context of image encryption, CNNs offer a unique approach by learning to encode and decode images directly from their pixel values, thereby circumventing the need for explicit algorithmic rules [6].

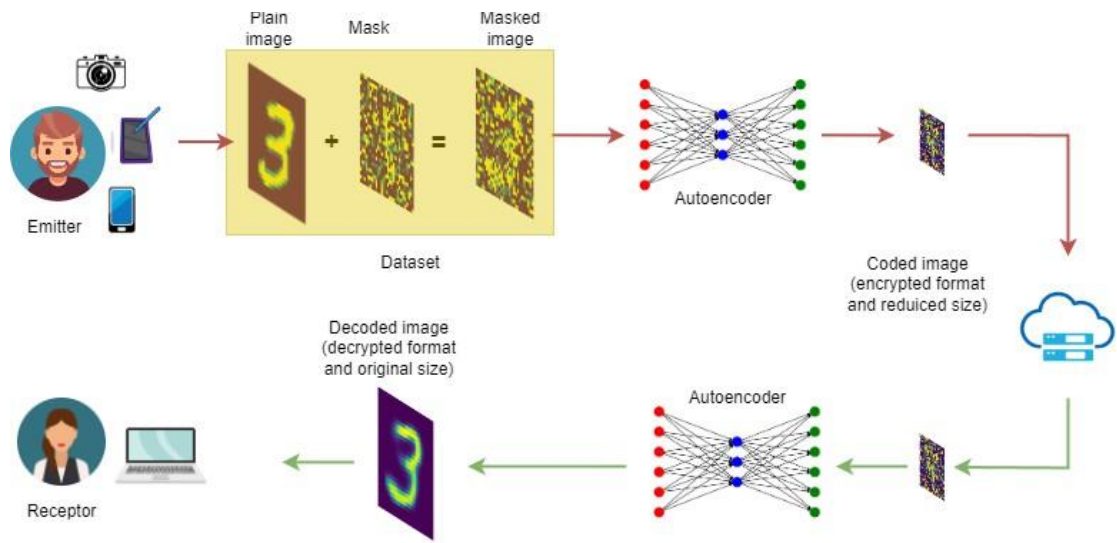

Figure 1.Deep Learning image protection model

In this work, we extend our work [7] initially presented at the *8th International Conference on Networks and Security (NSEC2024)*, we delve into the implementation and evaluation of a DL-based system for image protection and compression. The principle consists of using a masked dataset(CIFAR-10 and MNIST) to train the designed autoencoder. Webegin by generating the masked dataset by combining the original images with a randommask. This mask is used as the secret key established between the authenticated parts that have authorized access to the data. This mask can also be used by the receiver client to recover the plain image. Then, we designed and trained an autoencoder to learn and protect the plain image using the masked dataset and send the encrypted image. The receiver can decrypt the image in a symmetric way using the same trained model. The method allows data privacy protection. The proposed method demonstrated its effectiveness in preserving data integrity in addition to reducing the dimension of the sent encrypted image, from 2352 Bytes to 784 Bytes on the MNIST dataset, and from 3072 Bytes to 1024 Bytes on the CIFAR-10 dataset. Therefore, a unique model can serve as a model for encryption and decryption in applications based on symmetric cryptography models and can be used for image compression applications. The whole architecture of a practical example of the use of this method is illustrated in Figure 1.

By analyzing the performance, security level, internal architecture, and computational complexity of classic encryption techniques, we aim to provide valuable insights into the efficacy and trade-offs inherent in modern image encryption techniques [8]. The security analysis of the obtained results proves that the proposed model is promising for the new generations of image encryption applications.

The remainder of this paper is organized as follows. Section 2 discusses the wholer elated work, and Section 3 presents the used datasets, the generated masked data, and the internal architecture of the designed method in addition to the implementation of the processing steps of the proposed autoencoder in both the encryption and decryption phases. Section 4 presents and discusses the experimental results and gives a brief security analysis of the proposed model.

Finally, Section5 concludes this work and gives some directions for our future work to improve the limitations of the trained model.

## 2. RELATED WORK

As we know, in the last decade, machine learning and artificial intelligence have emerged inmanydomainsofourdailylivessuchascomputervision,objectdetection,e-health,smartcity,

Smart homes, smart cars, etc., but also in data protection and cryptographic applications. In this section, we expose some related works that used AI techniques in the area of preserving information security applied principally for data privacy protection and digital data encryption.

- A Deep Learning-based Stream Cipher Generator for Medical Image Encryption and Decryption: DeepKeyGen by Yi Ding et al. [9], is a novel deep learning-based key generation network for encrypting and decrypting medical images. By employing a generative adversarial network (GAN), they aim to generate a private key, with a transformation domain guiding the learning process. DeepKeyGen seeks to learn the mapping relationship between initial images and private keys. Their evaluation of three datasets, including the Montgomery County chest X-raydataset, the Ultrasonic Brachial Plexus dataset, and the BraTS18 dataset, demonstrates the network's capability to achieve high-level security in key generation.
- Image to Perturbation: An Image Transformation Network for Generating Visually Protected Images for Privacy-Preserving Deep Neural Networks: Made by Hiroki Ito et al. [10], it introduces an image transformation network aimed at generating visually protected images for privacy-preserving deep neural networks (DNNs). Unlike conventional perceptual encryption methods, this network maintains image classification accuracy while exhibiting strong robustness against various attacks, including DNN-based ones. The absence of security keys further simplifies the process. Experimental validation showcases the network's capability to protect visual information while preserving high classification accuracy using ResNet and VGG classification networks. Additionally, the visually protected images demonstrate resilience against diverse attacks, affirming the efficacy of the proposed transformation network in ensuring privacy in DNN applications.
- Learnable image encryption: In recent years, many researchers have explored the existing learnable image encryption schemes and proposed new ones. Among others, we can cite Tanaka [11] who presented the state-of-the-art privacy-preserving deep neural networks and proposed a shame based on encrypting images to will be not recognized by the human eye but still learnable by analysis with a machine, Sirichotedumrong et al. [12, 13] who presented another scheme (named as SKKscheme) using independent encryption keys unlike the basic Tanaka scheme using only one key, recently Sirichotedumrong et al. [14] proposed an image transformation scheme based onGANs, proving that the need to manage encryption keys no longer existed, and Huang [15] et al. proposed a learnable image encryption scheme that is an enhanced version of previous methods and can be used to train a great DNN model and simultaneously keep the privacy of training images.
- Image encryption based on autoencoders: With the advancement of the AI techniques, use of hybrid cryptographic algorithms has formed a new generation of cryptosystems. For example, Y. Sanget al. [16]proposed a novel image encryption method based on logistic chaotic systems and deep autoencoder. In the encryption phase, they randomly scrambled the plaintext image using a logistic chaotic system. Then, they encoded this image by a deep autoencoder to generate the ciphered one. A. Fawad et al. [17]

proposed a new encryption scheme for colour images employing a convolutional autoencoder for dimensionality conversion, DNA and chaos to perform the encryption/decryption phases. B.Wang et al. [18] proposed a new image compression and encryption framework that integrates encryption algorithms with a deep-learning compression network employing an autoencoder. For a higher level of visual security, they replaced the parameters of the synthesis network with a new parameter matrix based on a logistic map controlled by a secret key.

Note that the discussed works in this section are based only on models designed to ensure image protection and data privacy. However, our proposed model utilizes a convolutional autoencoder and masked datasets for image encryption and decryption, demonstrating efficacy in preserving data integrity while reducing dimensionality. Therefore, the proposed model can be used on onehand as a visual image encryption algorithm, and on the other hand as an image compression algorithm. Based on this analysis, we can conclude that our unique model can serve as a solution for two different problems of image processing applications.

## 3. PROPOSED AUTO ENCODER ARCHITECTURE

In addition to the main use of CNN in computer vision applications (object detection, tracking car traffic, people recognition, segmentation…), in the last years, many works based on CNN have been published in the field of image encryption [15, 19, 20, 21]. In this paper, we propose a visual image encryption and compression model that can be used to protect data privacy and compress images.

The designed model is an autoencoder using convolution functions based on filters to detect patterns, edges, shapes, and colors from original images, maxpooling function to reduce the dimension of the images before transmission, upsampling functions to recover the original dimension of the images after reception, RELU activation functions to increase the non-linearity of the generated outputs, bitwise XOR operation to combine plain images with the key mask to generate the protected images. After several rounds of training through the forward and backpropagation protocol, where it learns to minimize and adapt its weight and biases to approach the original and expected images. As a result, the used model allows for learning complex patterns from plaintext and masked data to generate acceptable ciphertext in reduced dimension or to reconstruct plain text from features of the ciphertext and recover the plain image in the original dimension. Therefore, the global model known as autoencoder is formed by two main blocks, an encoder to generate ciphertext, and a decoder to reconstruct the plaintext.

Before the encryption process, we begin by combining a bitwise XOR operation between the plain images and a random mask that can be considered as the secret key established between the emitter and the receptor. These masked images are then grouped and added to anew masked dataset that will be used to train, test, and validate the model. After that, the autoencoder is designed based on implementing multiple layers, including convolutional layers followed by a RELU activation function, max-pooling layers, and the neural link or fully connected layers, the features of the input masked image are extracted and then used to produce the output cipher image in a reduced dimension. The details of each used layer in the architecture are given in Figure 3. This step is known as the encoder phase.

The decryption process is based on implementing the same network layers in reverse order. Therefore, the autoencoder receives the encoded image and applies multiple layers, including convolutional layers followed by a RELU activation function, and upsampling layers to recover the original dimension of the image contrary to the pooling function that reduces the dimension(see Figure 2 for an explanation), and the neural link or fully connected layers, the features of the coded image are extracted and then used to produce the recovered image (as similar as possible to the plain image). This step is known as the decoder phase, and the details of each used layerin the architecture are given in Figure 3.

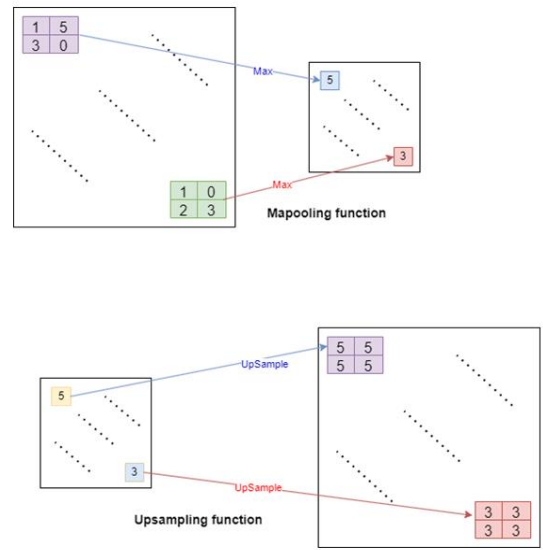

Figure 2. Comparison between Maxpooling and Ups ampling functions

After designing the neural network model firstly, it was trained, tested, and validated secondly using the MNIST and CIFAR-10 datasets. During the training step, the autoencoder learns to encrypt the important information of a masked image in a latent space of reduced imension and then reconstruct the image from this latent space. To reconstruct the plain image from the visually encrypted and masked one, we minimized the difference between both images by measuring the loss function (MSE loss) with a learning rate fixed to 0.001. As a result, the recovered image is an approximation of the original image, based on the information contained in its encoding. The performance of the proposed model is better on the MNIST dataset compared to the CIFAR-10 dataset. Examples of processed images from each dataset through the autoencoder are given in Figures 4 and 5 where the visually encrypted and masked image is generated after processing of the encoder phase, and the reconstructed image is generated directly unmasked after processing the decoder phase. As we can remark the MNIST decoded image is close to the input plain image, but the CIFAR-10 decoded image is still noised and requires more processing steps to recover the expected plain image. This is the limitation of this model in this dataset. We discuss in detail the advantages and limitations in the next section.

## 4. EXPERIMENTAL RESULTS AND DISCUSSION

In this section, we present and discuss the experimental results generated by the proposed autoencoder applied in both the MNIST and the CIFAR-10 datasets. We also analyze security properties, the advantages, and the limitations of the proposed model.

### 4.1. Used Datasets

In this work, we used two datasets, the CIFAR-10 database which contains 60.000 color images

117

of size 32×32 pixels divided into 10 classes (Airplane, automobile, bird, cat, deer, dog, frog, horse, hip, truck), and every class contains 6000 images. The global space is divided into a training space which contains 50.000 images (5000 images from each class), and a test space which contains 10.000 images (1000 images from each class). Similarly, we used also the MNIST database which contains 60.000 black and white images of size 28×28 pixels divided into 10 classes of handwritten digits (0, 1, 2, 3, 4, 5, 6, 7, 8, 9), and every class contains 6000 images. The global space is divided into a training space which contains 50.000 images (5000 images from each class), and a test space which contains 10.000 images (1000 images fromeach class).

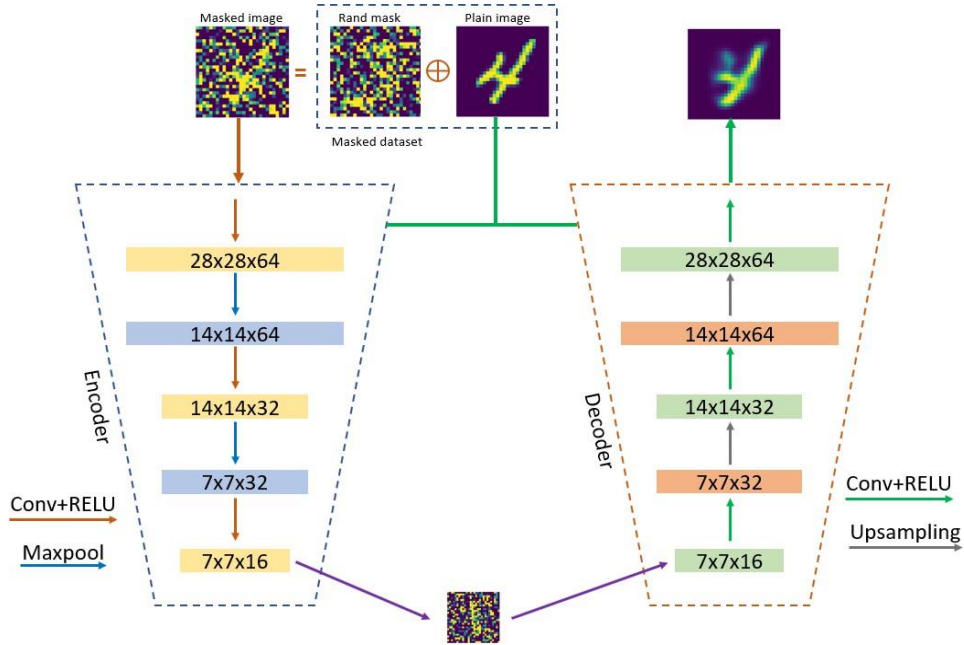

Figure 3. The internal layers of the proposed autoencoder

## 4.2. Experimental Results and Discussion

After training the model, it is tested and validated in both the MNIST and CIFAR-10 datasets. The experimental results of each of them are shown in Figures 4 and 6, respectively.

Firstly, analyzing the generated images by the proposed architecture on the MNIST dataset (see Figure 4), it is remarkable that the generated images by the autoencoder are visually encrypted (see Figures 4 (b), (e), (h), (k), and (n)) and it is difficult to recognize the plain data without decrypting it. Similarly, we remark that the generated images on the CIFAR-10 dataset are also well encrypted (see Figures 5 (b), (e), (h), (k), and (n)). This data can be sent over the existing networks in this protected format, and even if an attacker captures this data, he will not be able to decrypt it without knowing the used trained model and the mask applied to the original image. Unlike plain images which have a localized distribution of pixels, the encrypted images have a distribution close to the uniformone. Therefore, this architecture guarantees data privacy protection as expected by the users of the cyber digital world. After receiving and decrypting the data using the trained model, the recovered images from the MNIST dataset are close to the original plain ones (see Figures 4 (c), (f), (i), (l), and (o)). However, the recovered images from the CIFAR-10 dataset are noisy (see Figures 5 (c), (f), (i), (l), and (o)). Actually, this is a limitation of this model that we will expect to improve in future works.

Secondly, considering the mask used to visually protect the images before encoding them as a secret key, the resistance of the model against brute force attacks will be improved and will surpass all standard algorithms using a fixed secret key of size 128 bits to 512 bits and keyspace of $2^{128}$ to $2^{512}$. Therefore, for example for a mask of the same size as the CIFAR-10 images, $32\times32\times3\times8$, the key space of the proposed model can reach $2^{24576}$, which makes brute force attacks theoretically impossible.

Thirdly, using this autoencoder to transmit digital images in a reduced dimension can make data transfer 3 times faster. For example, to transmit 25 images of size $32\times32\times3\times8$ bits using a processing unit of 100 Mbps, we need 0.6144 µs. However, the generated image using the proposed architecture will be $8\times8\times16\times8$ bits. Therefore, to transmit the same 25 images using the same processing unit, we need 2048 µs. As we can remark, we reached an acceleration of 3 times.

Finally, we conclude that experimental results show that the proposed CNN-based model exhibits proficient encryption and decryption processes on the MNIST dataset, and proficient encryption process and acceptable decryption process on the CIFAR dataset, despite some noise in the recovered images. So, it demonstrated a notable strength in image visual encryption. This discrepancy highlights the potential of deep learning models in image encryption and data privacy protection applications. Despite its shortcomings in decryption, the model's success in encryption underscores its promise in data security. Further refinement of decryption capabilities within deep learning models is necessary to fully leverage their potential in robust encryption tasks. In addition, the proposed autoencoder can be useful, on one hand, to visually encrypt and protect images, and on another hand to compress and reconstruct images.

## 5. CONCLUSIONS

In this study, we conducted a comprehensive analysis to guarantee data privacy protection for digital images using DL and CNN-based techniques. To achieve this objective, we first created a masked images-based dataset using the publicly known MNIST and CIFAR-10 datasets. The principle was to combine the plain images with a random mask using a bitwise XOR operation. Then we used this new dataset to train, test, and validate our model. The proposed architecture is based on an autoencoder that is able to mask and encode an input plain image to generate an encrypted one that will be sent ina protected format. On the reception side, the same trained model will be used to decode and recover the original plain image. In addition, the images are sent in reduced dimensions which can accelerate 3 times the data transfer speed. We also showed the good security properties of the proposed model by analyzing the distribution of the generated close to the uniform distribution and the key space of $2^{24576}$, which makes brute force attacks theoretically impossible.

Therefore, we conclude that the propped deep learning model is promising and can be used in two different applications of image processing, either for image encryption and privacy protection or for image compression.

Despite its limitations in decrypting images from the CIFAR-10 dataset, the proposed method excelled in the encryption of digital images in two datasets, namely MNIST and CIFAR-10, highlighting its potential in security applications.

In our future works, to address the encountered limitation of the decryption phase, we explore and adapt architecture for the CIFAR-10 dataset based on a deep model using more internal layers. We expect also to use more sophisticated masks generated by a robust random-number generator, like chaotic maps (logistic map, skew-tent, piecewise linear chaotic map…) to enhance the visual protection before encoding and extending the constructed masked dataset. Additionally, we explore implementing the proposed model on a hardware FPGA-based platform to evaluate its performance and its suitability for real-time applications.

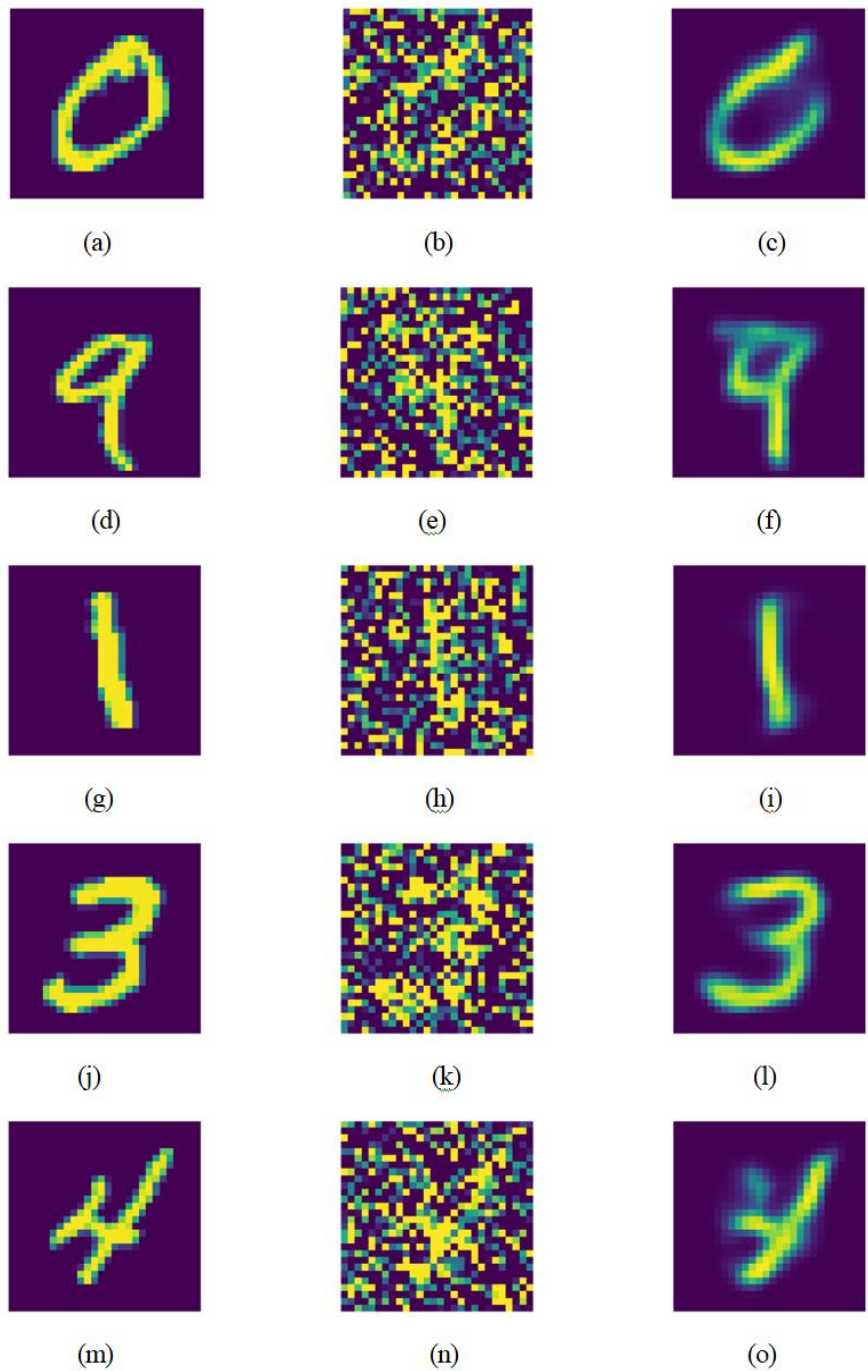

Figure 4. The experimental results for encrypted and decrypted images from the MNIST dataset

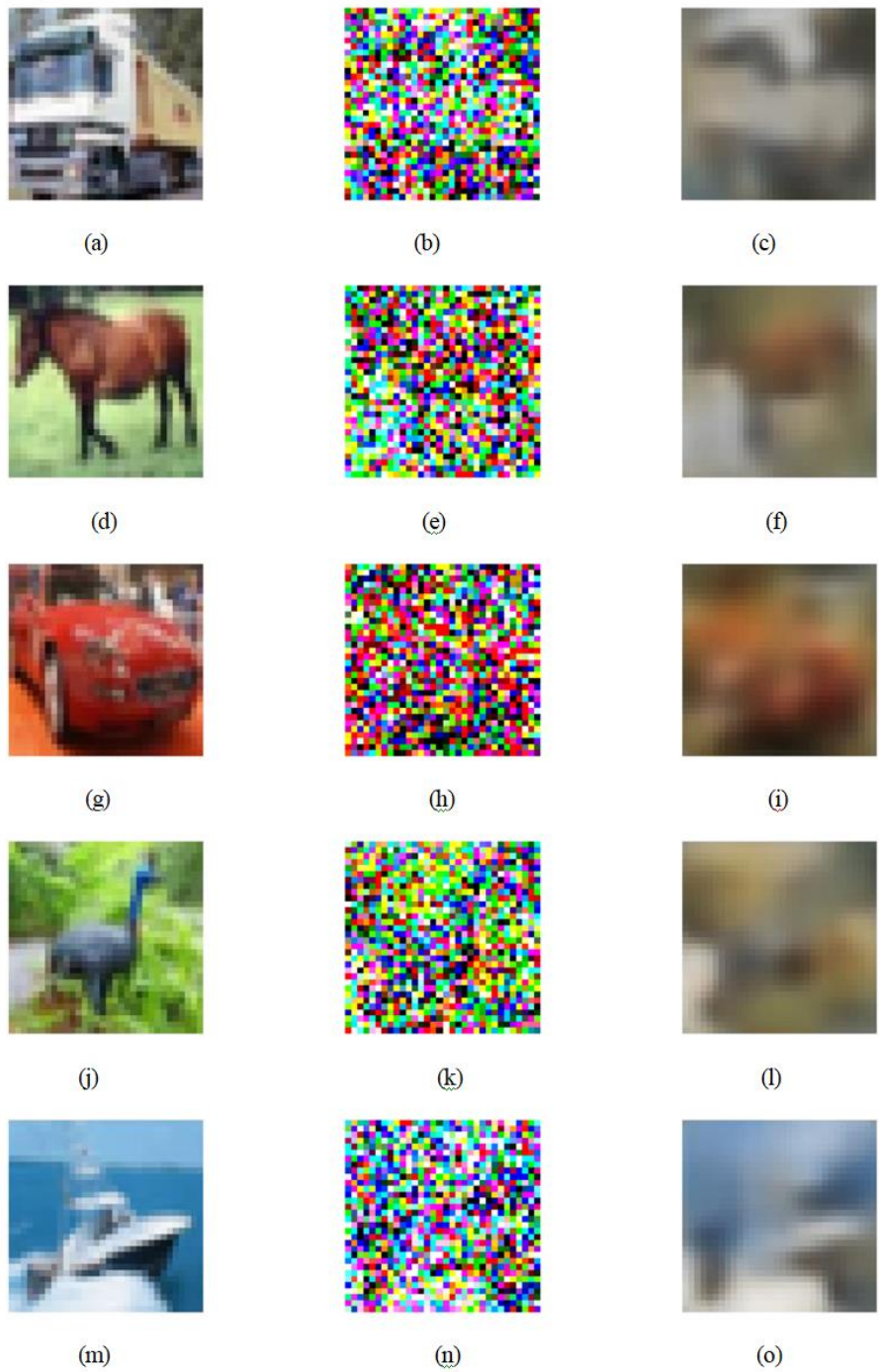

Figure5.The experimental results for encrypted and decrypted images from the CIFAR dataset

## CONFLICT OF INTEREST

The authors declare no conflict of interest.

## ACKNOWLEDGMENT

This work was supported by the Bourgogne Franche-Comte region as part of the ANER number2024PRE00022project entitled CIAPD.

## AUTHORS

**Mahdi Madani** is an associate professor at the University of Burgundy, Image et Vision Artificielle (Imvia) laboratory. His main research interests are information security, design, and hardware implementation of algorithms/architectures, deep-learning for image protection applications. He received his Ph.D. degree in Electronics Systems from the University of Lorraine in July 2018. He was temporary research and teaching associate at IUT Auxerre (2 years), and IUT Nantes (2 years).

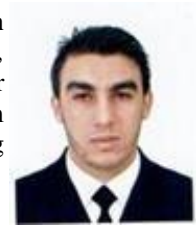

**El-Bay Bourennane** is currently a Professor of Electronics with the Laboratory of Image et Vision Artificielle (ImVia), University of Burgundy, Dijon, France. His research interests include dynamic reconfigurable system, image processing, embedded systems, and FPGA design and real-time implementation.

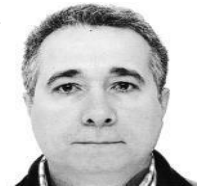